\documentclass{article}
\usepackage{aip}

\input{tcilatex}

\begin{document}

\title{Slow dynamics and correlation functions }
\author{V. Halpern \\
Department of Physics, Bar-Ilan University, Ramat-Gan 52900, Israel}
\maketitle

\begin{abstract}
The slow dynamics of a system as it approaches a phase transition,
associated with the slowing down in the decay of a correlation function, can
be caused by a sharp increase in the probability of a particle's returning
to its original state following a transition, rather than to a slowing down
in the transition rates as is usually assumed. The results of our
calculations show that this is the case for the ferromagnetic Potts model.
The implications of this result for various theories of the glass transition
are discussed.
\end{abstract}

\section{Introduction}

The drastic slowing down of a system's dynamics as the temperature $T$ is
lowered towards a critical temperature $T_{c}$ is a well-known feature of
many physical systems. Typical examples include the slowing down of the $%
\alpha $-relaxation process in supercooled liquids as the glass transition
temperature is approached and the critical slowing down of the dynamical
processes in a system undergoing a phase transition as the phase transition
temperature is approached. There are two aspects of this slowing down,
namely the general theory and the nature of the specific mechanism. For the
general theory, in the case of phase transitions this slowing down can
readily be derived quite simply from the Landau theory of phase transitions,
which involves an expression for the free energy of the system in terms of
an order parameter and its fluctuations.\cite{Phasetrans}, while for the
glass transition there are various suggestions for general theories but none
that is yet generally accepted \cite{glasstrans}. With regard to the nature
of the specific mechanism, it is usually assumed that this slowing down is
associated with a drastic slowing down in the rates at which the particles
make transitions. In this paper, we show that there is another possible
mechanism, namely a change in the results of successive transitions, and
present results for a specific model system where this is the case.

The calculation of a system's properties is usually performed in terms of
correlation functions $C(t)$, that describe the average relationship between
the states of the particles at time $t$ with their original state at time $%
t=0$. One reason for this is that, according to Kubo's formula, many of a
system's observable macroscopic properties are determined by correlation
functions. For instance, the frequency dependent dielectric susceptibility $%
\chi (\omega )$ is proportional to the Fourier transform of the
dipole-dipole correlation function $<M(t).M(0)>$ \cite{diel}. Hence, a rapid
increase in the probability that molecules return to their original state
(in the relevant phase space) following a transition, which we call the
return probability, will also lead to a drastic slowing down in the rate of
decay of the correlation functions, and so of the associated relaxation
processes and of the system's dynamics. The question of whether this
process, rather than a drastic slowing down of the transition rates, is
responsible for the slow dynamics is well worth investigating. For instance,
for supercooled liquids it has important implications for the type of theory
that can explain the glass transition, as we discuss below. Incidentally,
this type of approach is in line with that of Berthier and Garrahan \cite%
{Berthier}, who emphasized the importance of considering transitions in real
space.

The property of a system that is most often studied, both theoretically and
experimentally, is the correlation function, but this on its own cannot show
which of the above mechanisms is responsible for the slowing down. Another
property that can readily be studied theoretically (although not so easily
measured experimentally), and which can throw some light on this question,
is the fraction of particles in the system that have never changed their
state up to time $t$. We call this function the system's transition function
and denote it by $Pt(t)$. If the particles move randomly after making a
transition, the correlation and transition functions should decay with time
in a very similar manner. However, if the particles have a tendency to
return to their original state because of the arrangement of their
neighbors, as in Funke's mismatch and relaxation model for\ ionic conductors %
\cite{cage} and in the cage effect in mode coupling theory \cite{MCT-cage}\
for instance, the correlation function will decay much more slowly with time
than the transition function. It follows that a comparison of the change in
the time dependence of these functions as the critical temperature is
approached can throw light on which of the above mechanisms is responsible
for the slow dynamics. In this paper, we make such a comparison for a simple
model system in which there is a phase transition, and so critical slowing
down as it is approached, and then consider the implications of our results
for other systems.

The system that we consider is the ferromagnetic $q$-state Potts model, for
which we have recently reported results for the transition function $Pt(t)$,
in a paper \cite{Halpern-JCP} henceforth referred to as I. The function $%
Pt(t)$ was there called $P(t)$ and termed the relaxation function, but we
have changed its name and symbol in this paper so as to avoid any impression
that it is related to a physical property such as the dielectric
polarization. We have extended the calculations in I by calculating
simultaneously the transition function $Pt(t)$ and the normalized
correlation function $Cr(t)$, which we define below, as $T_{c}/T$ increases
towards unity. In section 2, we describe briefly the model system and how we
analyzed its properties. The results of our calculations are presented and
discussed in section 3, while our conclusions are summarized in section 4.

\section{The Potts model and its analysis.}

The Hamiltonian for the ordered ferromagnetic $q$-spin Potts model with
interactions only between the spins at adjacent sites can be written as \cite%
{Wu-rev} 
\begin{equation}
H=-J\sum_{i}\sum_{j(i)}\delta (\sigma _{i},\sigma _{j})
\end{equation}%
where $J>0,$ the first sum is over all the sites $i$ in the system and the
second one over all the sites $j(i)$\ that are nearest neighbors of the site 
$i$, the spins $\sigma _{i}$ can take any integer value between $1$ and $q$,
and $\delta $ is the Kronecker delta, $\delta (a,b)=1$ if $a=b,\quad 0$ if $%
a\neq b$. For this system, some care is needed in defining the relevant
correlation function. The elementary correlation function is just $%
s(t)=<\sum_{i}\delta \lbrack \sigma _{i}(t),\sigma _{i}(0)]>$, but this will
never tend to zero as $t\rightarrow \infty $ since the spins can only assume 
$q$ distinct values, so that even for completely random $\sigma _{i}(t)$ a
fraction $1/q$ of them will equal $\sigma _{i}(0)$. Accordingly, for a
system of $N$ spins we define the normalized correlation function $Cr(t)$ by 
\begin{equation}
Cr(t)=<\sum_{i}\{\delta \lbrack \sigma _{i}(t),\sigma
_{i}(0)]-1/q\}>/[N(1-1/q)]
\end{equation}%
for which $Cr(0)=1$ and if eventually there is no correlation between the
initial and final states then $Cr(\infty )=0$.

It follows from equation (1) that the energy $E$ associated with a site on
which the spin is equal to that on $z$ adjacent sites is just $-zJ$. The
probability of a change in the spin at a site which involves an increase of $%
\Delta E$ in this energy at temperature $T$ was taken to have the standard
form 
\begin{eqnarray}
w &=&w_{0},\quad \Delta E<0 \\
w &=&w_{0}\exp (-\Delta E/k_{B}T),\quad \Delta E>0  \nonumber
\end{eqnarray}%
For an infinite square lattice, the values of the critical temperature $%
T_{c} $ are given by the solutions $v=v_{c}$ of the equation $v^{2}=q$,
where $v=\exp (J/[k_{B}T])-1$, while the phase transition is of second order
for $q\leq 4$ and of first order for $q>4$ \cite{Baxter}.\ In order to
examine both types of phase transition, we performed our calculations for
two systems, with values of $q=3,6$. For our simulations we considered, as
in I, a square lattice of 200 x 200 sites, and we here chose $w_{0}=1$ since
we are interested in temperatures close to $T_{c}$. The simulation
techniques used were the same as in I, apart from starting the anneals from
an initial state of all identical spins instead of one with random spins.
Once a steady state was reached, five sets of several successive simulation
runs were performed on the system, with each run proceeding until the spin
had changed at least once at 99.9\% of the sites. A maximum of 5 runs was
performed in each set, but these were stopped once $Cr(t)$ became less than $%
0.002$. Finally, the average values of $Pt(t)$ and $Cr(t)$ for the five sets
of runs were calculated.

The results were analyzed in terms of the mean relaxation times, $<\tau >$,
for the decay of these functions, and two different techniques were used to
calculate them. In general, a monotonically decreasing function $f(t)$ can
be expressed as a sum or integral of functions $\exp (-t/\tau )$, with a
distribution $g(\tau )$ of relaxation times $\tau $, $f(t)=\int g(\tau )\exp
(-t/\tau )d\tau $. In that case, $<\tau >=\int g(\tau )\tau d\tau
=\int_{0}^{\infty }f(t)dt$, and this latter expression was used to calculate 
$<\tau >$. The main method that we used for our calculations was to fit $%
Pt(t)$ and $Cr(t)$, for values of these functions greater than 0.01, to
stretched exponential functions $f(t)=A\exp [-(t/\tau _{0})^{\beta }]$, for
which $\int_{0}^{\infty }f(t)dt=(A/\beta )\Gamma (1/\beta )\tau _{0}$. In
all cases, an excellent fit was obtained (in a graph of $\ln [f(t)]$ as a
function of $t$) with $A=1$. We restricted our calculations to values of $%
x=T_{c}/T$ such that after a set of 5 runs $Cr(t)<0.1$, since for higher
final values of $Cr(t)$ it is difficult to justify a fit to such a function.
In addition, $<\tau >$ was calculated directly for the two functions from $%
\int_{0}^{\infty }f(t)dt$, and the value obtained was found to be very close
to that obtained from the fit to the stretched exponential function (except
for $Cr(t)$ at the highest values of $x$, where the cut-off in $Cr(t)$ after
five runs led to a much lower value for $\int_{0}f(t)dt$ since the integral
did not extend to negligible values of $Cr(t)$ ). Accordingly, all\ the
results for $<\tau >$ that we present, which are denoted by $\tau _{P}$ for $%
Pt(t)$ and by $\tau _{C}$ for $Cr(t)$,\ are those derived from fits to a
stretched exponential function.

\section{Results and Discussion.}

In figure 1, we show both $\ln (\tau _{P})$ and $\ln (\tau _{C})$ as
functions of $T_{c}/T$ for $q=3$ and for $q=6$. Incidentally, the results
presented here for $\ln (\tau _{P})$ differ slightly from those in our
previous paper I because there we showed the stretched exponential
relaxation time $\tau _{0}$ rather than $<\tau >$, and also used $w_{0}=0.5$%
. It is immediately apparent that while for $T_{c}/T=0.85$ the relaxation
times $\tau _{C}$ and $\tau _{P}$ are similar, the relaxation times $\tau
_{C}$ for the correlation function becomes larger than those for the
transition function $\tau _{P}$ as $T_{c}/T$ increases, until for $%
T_{c}/T=0.99$ we find that $\tau _{C}/\tau _{P}$ $=33$ for $q=3$ and $6.3$
for $q=6$.\ In order to show the difference between the functions $Cr(t)$
and $Pt(t)$, and not just the ratio of their relaxation times, we present
these functions for $T_{c}/T=0.85$ and $T_{c}/T=0.99$ for $q=3$ in figure 2
and for $q=6$ in figure 3.\ These figures show clearly that for $%
T_{c}/T=0.85 $ the transition and correlation functions decay with time in a
very similar manner, but for $T_{c}/T=0.99$ the correlation function decays
much more slowly than the transition function. We also note from figure 1
that as $T_{c}/T\rightarrow 1$ while $\ln (\tau _{P})$ increases fairly
steadily, with no particular sign of a phase transition, $\ln (\tau _{C})$
starts to increase very rapidly, so that for higher values of $T_{c}/T$ than
those shown in the figure it decays very little on the time scale of around $%
5\tau _{P}$ to which we extended our simulations. Thus, it is the rapid
increase in $\tau _{C}$ (rather than in $\tau _{P}$) which is a sign of the
critical slowing down associated with a phase transition in our system, and
this is due to the rapid increase in the return probability.

For the Potts model, there is a simple qualitative explanation of this
increase in the return probability as the temperature decreases, and also of
the difference between the systems with $q=3$ and those with $q=6$. For a
site within a cluster of identical spins on a square lattice, the activation
energy $\Delta E$ required for a transition is $4J$. As a result, the
slowing down of the decay of the transition function, which is associated
with the temperature dependence of the factor $\exp (-\Delta E/k_{B}T)$\ in
equation (3) and with the increasing fraction of sites within clusters (as
found in I) as $T$ decreases, does not change dramatically as $T$ approches $%
T_{c}$. On the other hand, when the spin at a site within a cluster changes,
it can change again without any activation energy until it returns to its
original state, when it again requires an activation energy of $4J$ to
change it. Thus, the spins on sites within clusters have a strong tendency
to return to their original values unless the spins on the adjacent states
have changed before they do so, something that happens most easily at higher
temperatures, and also for small clusters since the spins at sites on the
boundary of a cluster require a lower activation energy to change. As a
result, the increase in the fraction of sites within clusters (and in the
mean cluster size) as $T$ decreases leads to a rapid increase in the return
probability, and so in the ratio of $\tau _{C}/\tau _{P}$, with a dramatic
increase as the transition to a single phase is approached. This is just an
example of critical slowing down as the phase transition is approached. It
sets in at lower temperatures (higher values of $T_{c}/T$) for $q=6$ than
for $q=3$ because the fraction of sites within clusters at a given value of $%
T_{c}/T$ decreases with increasing number of different possible values of
the spins, i.e. with increasing $q$, as was found in figure 8 of I.

The above results show that for the ferromagnetic Potts model the main
mechanism responsible for the critical slowing down as the phase transition
is approached is the increasing return probability and not a drastic slowing
down in the rate at which transitions occur. Such a mechanism is not
restricted to the Potts model, and not necessarily even to phase
transitions, and can well be relevant to the glass transition in supercooled
liquids. As we discussed in I, the Potts model can be regarded as a simple
model for plastic crystals, which have relaxation properties similar to
those of supercooled liquids \cite{plastcryst}. Accordingly, we now consider
the relevance of this type of process to the\ drastic slowing down of the
relaxation rates in supercooled liquids as the glass transition is
approached. Our results show that this need not be caused by a change in the
basic relaxation mechanism, and in fact no such change was found in recent
experiments by Huang and Richert \cite{Richert}. Instead, as for the Potts
model, it can be due to the local molecular arrangements being such that
even when a molecule instantaneously changes its position or orientation it
rapidly returns to its original lower energy state. Such a process is
obviously compatible with theories of the glass transition based on
heterogeneities \cite{heterog} and/or smeared out phase transitions \cite%
{glass-phasetrans}, since the Potts model exhibits a phase transition while
the clusters of states of identical spin in it are local heterogeneities. It
is also fully compatible with theories of the glass transition that involve
potential energy landscapes \cite{PEL}, for which the molecular transitions
correspond to jumps within a meta-basin and those which lead to diffusion of
the molecule (and so to decay of the correlation function) to jumps of the
system from one meta-basin to another. However, such a process does not fit
so well to theories based on free volume\ \cite{freevol}\ or on kinetic
constraints \cite{kincon}, where it is the basic transition rates of the
molecules that change drastically as the temperature is lowered. Of course,
it is quite possible that different mechanisms are the dominant ones in
different systems.

\section{Conclusions}

A comparison between the transition function $Pt(t)$ and the correlation
function $Cr(t)$ of a system that exhibits a drastic slowing down of its
dynamical properties can show whether this slow dynamics is caused mainly by
a slowing down in the transition rates of the particles or by an increase in
their tendency to return to their original state following such a
transition. For the ferromagnetic Potts model, the latter process was found
to be responsible for the critical slowing down as the phase transition was
approached. The question of which is the dominant process has important
consequences for the different theories of the glass transition in various
types of supercooled liquids.\

Captions for figures

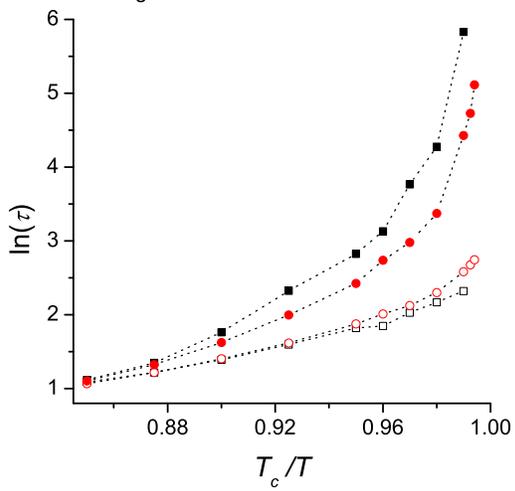
Figure 1. The mean relaxation time $\tau $ for the relaxation function $%
Pt(t) $ (open symbols) and for the correlation function $Cr(t)$ (closed
symbols) of the $q$-state Potts $\func{mod}$el as functions of $T_{c}/T$.
The results for $q=3$ are denoted by black squares, and those for $q=6$ by
red circles, while the broken lines are just to guide the eye.

Figure 2. The relaxation function $Pt(t)$ (continuous black lines) and the
correlation function $Cr(t)$ (broken red lines) for the 3-state Potts model,
for $T_{c}/T=0.85$ (lower pair) and for $T_{c}/T=0.99$ (upper pair).

Figure 3. The relaxation function $Pt(t)$ (continuous black lines) and the
correlation function $Cr(t)$ (broken red lines) for the 6-state Potts model,
for $T_{c}/T=0.85$ (lower pair) and for $T_{c}/T=0.99$ (upper pair).

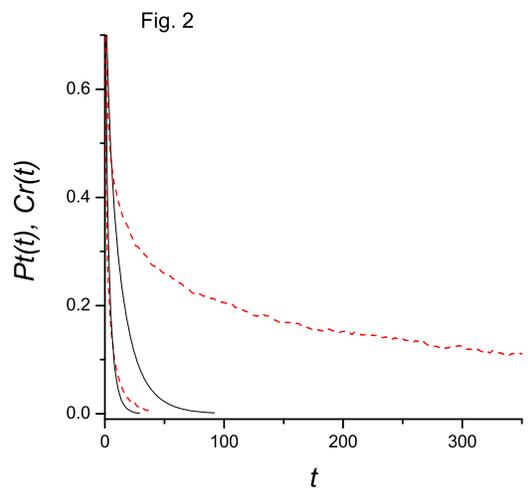
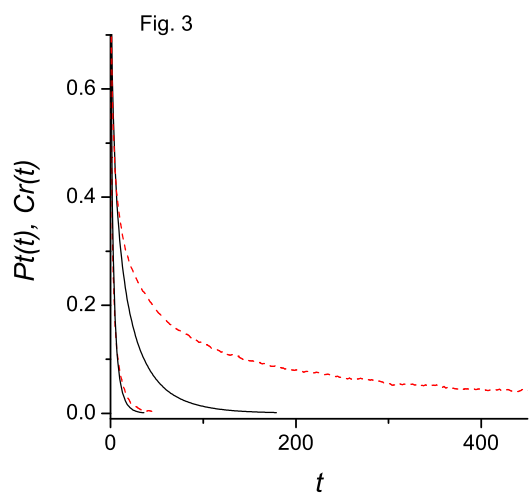
\end{document}